\documentclass[conference]{IEEEtran}
\IEEEoverridecommandlockouts
\usepackage{amsmath,amssymb,amsfonts}
\usepackage{algorithmic}
\usepackage{graphicx}
\usepackage{textcomp}
\usepackage[T2A,T1]{fontenc}
\usepackage{t1enc}
\usepackage[utf8]{inputenc}
\usepackage[english]{babel}
\usepackage{csquotes}
\usepackage{balance}
\usepackage{amsthm}
\usepackage{multicol}
\usepackage{nicefrac}
\usepackage{dsfont}
\usepackage{dblfloatfix}
\usepackage[dvipsnames]{xcolor}
\DeclareMathOperator{\I}{I}
\DeclareMathOperator{\HH}{H}
\DeclareMathOperator{\DD}{D}

\def\iC{{\cal C}}

\def\iE{{\cal E}}

\def\iP{{\cal P}}

\def\iT{{\cal T}}

\def\iX{{\cal X}}
\def\iY{{\cal Y}}
\def\iZ{{\cal Z}}

\newcommand{\sumfrac}[2]{\genfrac{}{}{0pt}{}{#1}{#2}}

\newcommand{\irott}{\mathcal}

\newcommand{\vm}{\mathbf m}

\newcommand{\vx}{\mathbf x}

\newcommand{\vy}{\mathbf y}

\newcommand{\vS}{\mathbf S}

\newcommand{\vi}{\mathbf i}
\newcommand{\vj}{\mathbf j}

\newcommand{\vV}{\mathbf V}

\renewcommand{\P}{\mathrm P}
\theoremstyle{plain}

\newtheorem{Lem}{Lemma}

\theoremstyle{definition}
\newtheorem{Def}{Definition}

\theoremstyle{remark}
\newtheorem{Rem}{Remark}
\usepackage[style=ieee,backend=biber]{biblatex}
\renewbibmacro*{bbx:savehash}{} 
\def\BibTeX{{\rm B\kern-.05em{\sc i\kern-.025em b}\kern-.08em
    T\kern-.1667em\lower.7ex\hbox{E}\kern-.125emX}}
\begin{document}

\title{Trellis Code Error Exponent From Results for Asynchronous Multiple Access Channels\\
\thanks{National Research, Development and Innovation Office (Hungary) – NKFIH K120706 and KH129601}
}

\author{
\IEEEauthorblockN{Lóránt Farkas}
\IEEEauthorblockA{\textit{Department of Analysis} \\
\textit{Budapest University of Technology and Economics}\\
Budapest, Hungary \\
lfarkas@math.bme.hu}
}

\maketitle
\begin{abstract}
An asynchronous multiple access error exponent result implicates a new result for time invariant trellis codes of memory 1.
\end{abstract}

\section{Introduction}

Trellis codes for discrete memoryless channels (DMCs) represent a generalization of block codes. A good early reference is Forney \cite{Forney}. The error exponents in \cite{Forney} are derived for time varying trellis codes, pointing out that there had been no success in proving them for time-invariant ones. Also the recent work of Merhav \cite{MerhavTrellis}, addressing error exponent of typical random codes consider ensembles of time-varying trellis codes. 

For time-invariant convolutional codes Shulman and Feder \cite{ShulmanFederTimeInvariantCC} gave time invariant error exponent, in logic size, for the first error event. Their exponent tends to the same exponent as in \cite{Forney}, as the parallel levels in the logic tends to infinity. However, their proof technique apply randomly generated matrices that does not say anything about errors of recurring codewords. 

Error exponents for multiple access channels (MACs) with two non-synchronous senders have been derived in \cite{Hawaii,Hongkong,Paris}. For a complete version see \cite{CAMACEEarXiv}. A key result is that controlled asynchronism (the senders transmitting with a chosen delay) may admit a larger error exponent than synchronous transmission (for a brief heuristic explanation, see below). While in \cite{Hawaii,Hongkong} both sender were admitted to use multiple codebooks, in \cite{Paris} the same exponents were shown achievable with each sender using only one codebook. 

The above mentioned improvement of the exponent in the asynchronous case can be explained heuristically as follows. In the synchronous case there are 3 error types. The first type is the case where only the first transmitter's codeword decoded incorrectly. The second type of error is the case where only the second transmitter has error. The third type of error when both transmitters codewords are incorrectly decoded. In many cases the third type of error is dominant (namely, has the largest probability). For example, when the channel is symmetrical, both user has the same rate and the rates are approaching the dominant face of capacity region (for dominant face, see Rimoldi et al. \cite{Urbanke}), then typically, the third type of error is dominant. In the asynchronous case several different error patterns exist, see Figure \ref{FigSubblocks}. These patterns always have parts where only one transmitter is decoded incorrectly. Those part can help the decoder to detect an error.

To prove directly that the asynchronous exponent is larger
than its synchronous counterpart a special model was applied, as state of the art
sphere packing bounds (upper bounds for the best
possible error exponents) of synchronous multiple access
channels are prohibitively complex to numerically evaluate (see Nazari et al. \cite{Nazari}). The binary XOR channel whose output is sent through a binary Z-channel was analyzed (see, Figure \ref{FigModel}). This channel can be viewed as a single user Z-channel too, as a single user can take two codebooks, shift the codewords from one appropriately, take XOR and send the result through the Z-channel (the idea comes from Haim et al. \cite{Arunjavaslat}). It was found, that the asynchronous exponent of the composite channel is higher than the sphere packing bound of the single user Z-channel.

At first glance, this seems to be a contradiction (namely, a better
exponent than the upper bounds of the best possible one was given) but the sphere packing bound is for block coding only.  Proper analysis of the above described single user coding schema (with two codebooks, shift, etc.) reveals that this is a variant of Trellis coding with memory 1.

As pointed out in \cite{CAMACEEarXiv}, MAC codes for controlled asynchronous transmission are related to trellis codes for DMCs. In particular, MAC codes with multiple codebooks (resp. one codebook) for each sender are related to time varying (resp. time invariant) trellis codes. The technique of \cite{Paris} that allows us to use one codebook during the communication process indicate that, the error exponent in \cite{Forney} are achievable also with time-invariant trellis codes. Leaving this conjecture open, here we further elaborate on the relationship of trellis codes and MAC codes.

In Section \ref{SecKnownResult} after reviewing the notations, the known numerical values of achievable error exponent of time-variant trellis codes from \cite{Forney} and the known achievable asynchronous error exponent of \cite{Paris,CAMACEEarXiv} are presented for the aforementioned specific DMC, showing the asynchronous error exponent is larger, in the low rate region, than the exponent of \cite{Forney} of memory 1, and in the high rate region they are close to each other. However, our asynchronous code is not time invariant, only periodically time-invariant with period 2.

\setcounter{figure}{1}
\begin{figure*}[!bt]
\begin{center}
 \includegraphics[width=0.7\textwidth]{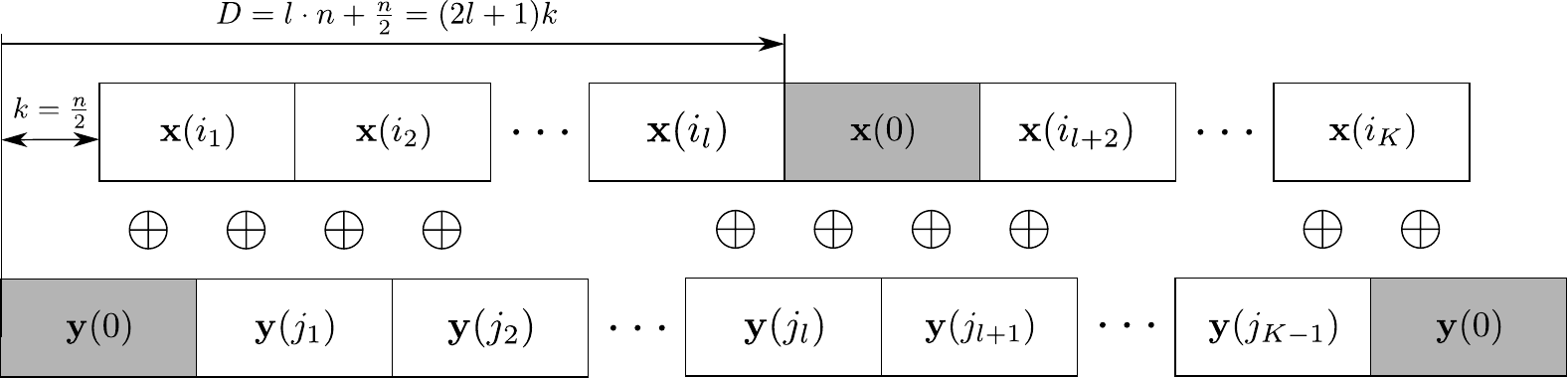}
\end{center}
\caption{ Combining the two virtual input streams to get a real input stream}\label{FigAsModel}
\end{figure*}

The result is improved in Section \ref{SecImproving} to full time invariant trellis code, using a generalization of the Packing lemma in \cite{Paris,CAMACEEarXiv}. The possibility of extending the results to general DMCs and systems with memory more than 1 is treated in Section \ref{SecGeneralization}.

\section{Comparing Known Results}\label{SecKnownResult}

\subsection{Notations, Definitions}

The set $\{1,2,\dots,l\}$ is denoted by $[l]$. Logarithms and exponentials are to the base 2 i.e., $\log x=\log_2 x$, $\exp x =2^x$. Polynomial factors will be denoted by $p_n$. Sequences will be denoted by boldface letter, so $\vx=(x_1, x_2, \dots, x_n)$

Random variables (RVs) are assumed to take values in finite sets. These sets, the RVs, and their possible values are typically denoted by calligraphic letters  and corresponding upper and lower case italics such as $\irott X,X,x$.

Distributions on finite sets are denoted by $P$ or more frequently by $V$, types/joint types ---empirical distribution--- of sequences are denoted by $P_\vx$. $P_{(\vx,\vy)}$, etc. The set of all distributions on $\iX$ and its subset of all types of sequences $\vx\in\iX^n$ are denoted by $\iP(\iX)$ and $\iP^n(\iX)$. Types (joint types) are frequently represented as (joint) distributions of dummy random variables, e.g. $P_{(\vx,\vy)}=V^{XY}
$.

\begin{Def}\label{DefTrellKod}
A trellis code with memory length $a$, rate $R$, blocklength $n$ is a codebook system that contains a codebook with $2^{nR}$ codewords of length $n$ for each time index $t$ and possible $(m_1, m_2, \dots, m_a)$, $m_i\in \{0\}\cup [2^{nR}]$ memory state consisting of $a$ previous messages. The trellis code is time invariant if these codebooks depend only on the memory state and not on $t$.  The initial memory state is  $(0, 0, \dots, 0)$ (here, 0 means no message), and after $K-a$ true messages 0-s are sent $a$ times, thus the final state is also $(0, 0, \dots, 0)$.
\end{Def}
\begin{Rem}
Note, that a classical block code is a trellis code with memory length $0$, so there is only 1 memory state ---the empty vector, and there is no leading and ending 0 sequences.
\end{Rem}

Error exponent achievable by rate $R$ trellis codes over a channel $W$ are given in \cite{Forney} as

\begin{align}
    P(\iE)\leq K_1\exp\left(-naE_0(\rho)\right) \label{EqTrellisHibaVsz}
\end{align}
where $E_0(\rho)$ is Gallager's function
\[E_0(\rho)=-\log_2\sum_{z\in \iZ}\left(\sum_{x\in \iX} p_x W(z|x)^{\frac{1}{1+\rho}}\right)^{1+\rho}\]
and $\rho$ is any parameter satisfying $0\leq \rho \leq 1$, $\rho<\rho_R$, where $\rho_R$ is the parameter that satifies $R=\frac{E_0(\rho_R)}{\rho_r}$.

\begin{Def}\label{DefCodeBooks}
 A MAC code of block length $n$ and rate pair $R_1,R_2$ is given by codebooks $\iC_1=\{\vx(i), i\in[2^{nR_1}]\}$, $\iC_2=\{\vy(j), j\in[2^{nR_2}]\}$ and in asynchronous case also by synch sequences $\vx(0)$, $\vy(0)$ and an integer $K\geq 2$. 
\end{Def}
In asynchronous case each sender transmits a synch sequence after $K-1$ consecutive codewords. The delay $D$, the numer of sybols between synch sequences, might be either unknown or chosen by the senders. As the synch sequences do not carry information, the effective transmission rates are $R_1\left(1-\frac{1}{K}\right), R_2\left(1-\frac{1}{K}\right)$.

\subsection{Model, known results}
We focus on the specific model depicted in Figure \ref{FigModel}, used in \cite{Paris,CAMACEEarXiv}. This model was used to prove that controlled asynchronism may outperform synchronism.\setcounter{figure}{0}
\begin{figure}[hbt]
\begin{center}
\includegraphics[width=0.25\textwidth]{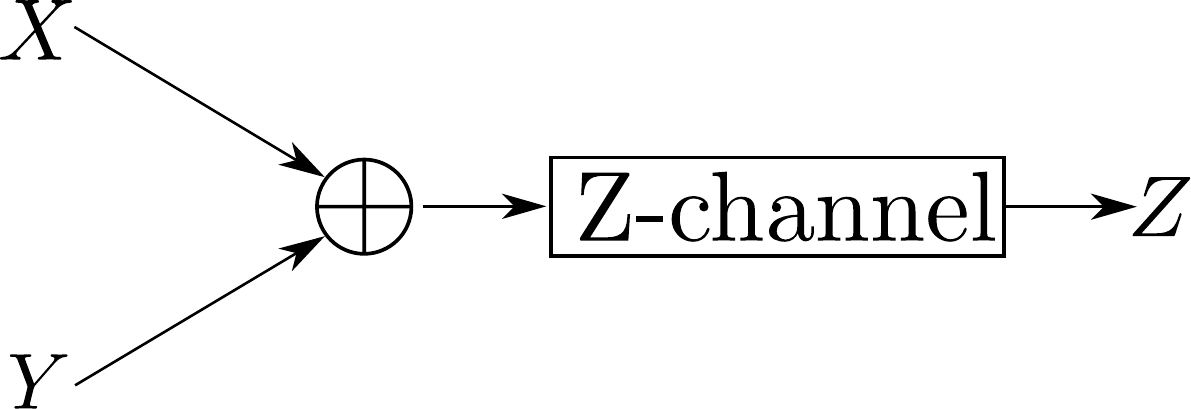}
\end{center}
\caption{ Multiuser channel for numerical calculations }\label{FigModel}
\end{figure}

This turns a (single user) Z-channel into a MAC with input alphabets $\iX=\iY=\{0,1\}$, where the XOR of the two inputs goes through the Z-channel. Formally, this MAC is defined by channel matrix 
\[ W(z|x,y)=W_z(z|x\oplus y), \quad x, y, z \in \{0,1\}\]
where $W_z$ is the matrix of the Z-channel.

Controlled asynchronous MAC codes are considered with $R_1=R_2=R$ for this virtual MAC that correspond to trellis codes of memory length 1 for the single user Z-channel. In this context the MAC is regarded virtual. The effective rate of this code is $R\left(1-\frac{1}{K}\right)$ same as the effective rate of the trellis code form Definition \ref{DefTrellKod} if $a=1$.

Suppose $n$ is even and $K$ is odd, $n=2k$, $K=2l+1$ and let the chosen delay $D=\left(l+\frac{1}{2}\right)n$. In a window of length $nK$, the MAC senders transmit concatenation, $\vx(i_1)\dots\vx(i_K)$, 
$\vy(j_1)\dots\vy(j_K)$ of codewords, with synch-sequence at the middle resp. end, i.e., $i_{l+1}=j_K=0$ (according to Figure \ref{FigAsModel}). More exactly, the second sequence is shifted circulantly by $k=\frac{n}{2}$, thus its transmission starts with the last $k$ bits and ends with the first $k$ bits of $\vy(0)$, see Figure \ref{FigAsModel}. All codewords and synch-sequence have type that tends to $P^*$, the capacity achieving distribution for the multi-access Z-channel.

The error probability $\P(\iE)$ of misdetecting any codeword is given by
\[\P(\iE)\leq p_n \exp(-n E_r(R)),\]
where $p_n$ is polynomial term in $n$ and the random coding error exponent $E_r(R)$ of the asynchronous multiple access error exponent in \cite{CAMACEEarXiv} is given in case of the above Z-channel by
\begin{align}
 &\min_{\sumfrac{L\in [K], V_1,V_{12}:}{V_1^X=V_1^Y=V_{12}^X=V_{12}^Y=P^*}} 
 \bigg[ \DD(V_1\|P)+\frac{L-1}{2}\DD(V_{12}\|P)\notag\\
 &+\left|\I_{V_1}(X \wedge Z |Y)+\frac{L-1}{2}\I_{V_{12}}(X\wedge Y\wedge Z)-LR\right|^+ \bigg] \label{EqAMACExpZnel}
\end{align}
where $P(x,y,z)=W(z|x,y)P^*(x)P^*(y)$ and 
\begin{align}
 \I_V(X\wedge Y\wedge Z)=&\HH_V(X)+\HH_V(Y)+\HH_V(Z)\notag\\
  &-\HH_V(X, Y, Z). \label{EqMultiInfDef}
\end{align}
and $\HH(\cdot)$ is the Shannon-entropy. For channels that are symmetrical in the inputs the exponent is almost the same, apart from the fixed type of constant composition codewords. For the general exponent for any MAC see \cite{CAMACEEarXiv}.

The maximizing error pattern is always an irreducible pattern (one ``continuous'' pattern, where only two subblocks are such where one of the transmitter has error, see \cite{Hongkong, Paris,CAMACEEarXiv}). So, the number of possibly maximizing patterns are linear in $K$. For every fix pattern the exponent is convex, so the minimizing distributions are the same for every $k$-length subblocks that has similar error type. The numerical calculations of the exponent for fix $R$ can be implemented by minimizing in the possible distribution triples for each irreducible pattern with different length, then taking the minimum of these numbers (for more on this, see \cite{Paris,CAMACEEarXiv}). 

An asynchronous code for the virtual MAC as above gives rise to a trellis code for the Z-channel with blocklength $k$ and memory $a=1$. Indeed, consider the interleaved message sequence $\vm=(i_1, j_1, i_2, j_2, \dots, i_K, j_K)$, thus at time $t=2s-1$ resp $2s$ the memory state is $j_{s-1}$ resp. $i_s$ and the actual message is $i_s$ resp. $j_s$. In other words, the $s$-th $k$-long block is depend only on $m_s, m_{s-1}$. The trellis codebooks, in this case, consists of XORs of initial and final $k$ bits of codewords from codebooks of Definition \ref{DefCodeBooks} for the virtual MAC. This trellis code is ``almost'' time-invariant, the codebooks in Definition \ref{DefTrellKod} depend on $t$ only through its parity.

Figure \ref{FigRExTrellEx} below shows numerical values of achievable error exponents for the trellis codes as above obtained from \eqref{EqAMACExpZnel}, and of those achievable by time-varying trellis codes according to \eqref{EqTrellisHibaVsz}. The cross probability for the Z-channel was $P_{1\to 0}=0.101$. The depicted rates are the single user effective rates $R$ and the exponents are calculated for blocklength $k$ (as the asynchronous multiple access random coding error exponent is in $n$ in \cite{Paris,CAMACEEarXiv}, those numerical values must be multiplied by 2, as well as the rates while we have $2^{nR}$ number of codeword that is $2^{k2R}$ number of codeword).

\setcounter{figure}{2}
\begin{figure}[h!]
\begin{center}
\includegraphics[width=0.4\textwidth]{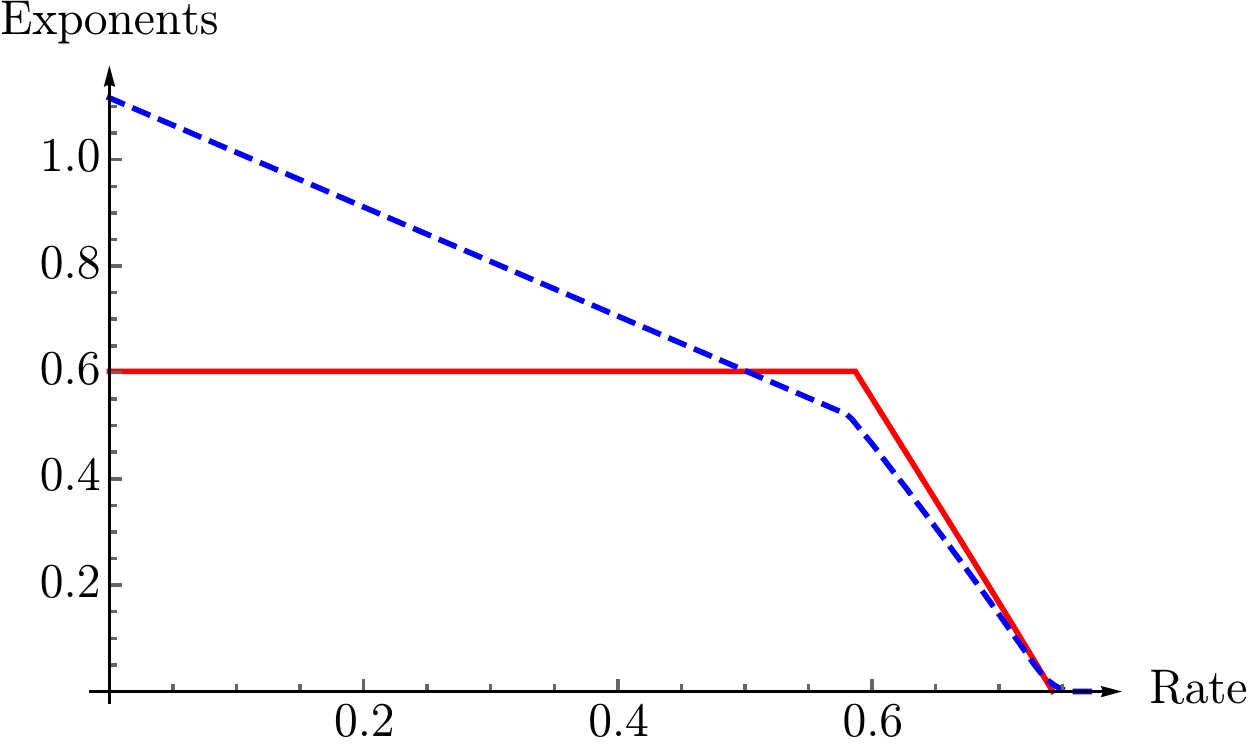}
\end{center}
\caption{ Numerical values of the asynchronous and the trellis exponent. Red continuous line depicts the lower bound of theoretically achievable time varying trellis exponent presented e.g. in \cite{Forney}, blue dashed line depicts the asynchronous MAC error exponent. }\label{FigRExTrellEx}
\end{figure}

Moreover, the asynchronous multiple access random coding error exponent in \cite{Hawaii,Hongkong, Paris,CAMACEEarXiv} is a universal exponent, so the decoding process does not depend on the channel matrix $W$.

\section{Improving the Result}\label{SecImproving}

The trellis code coming from asynchronous MAC in Section \ref{SecKnownResult} is not time invariant, while the two senders use different codebooks. Two 
codebooks are employed alternatingly, one which stands from the XOR of first (virtual) user codewords' first $k$ symbol, and second user codewords' last $k$ symbol and a second which stands from the XOR of first user codewords' last $k$ symbol, and second user codewords' first $k$ symbol. However, symmetry admits to prove that the two virtual codebook may be the same. With this, a time invariant trellis code will be obtained. Except the insertion of synch sequences with period $K$.

The Csiszár-type exponent \eqref{EqAMACExpZnel} is proven with the help of a Packing lemma which gives a bound for number of message pairs ---or in multiuser setting triplets, quadruples, etc.---in the codebooks with given joint type. This upper bound then ``plugged in'' to the error bound and applying standard method of types techniques gives the exponent. This Packing lemma is usually proven by random selection, similarly as the technique of Gallager, but any codebooks with the properties in the Packing lemma achieve the claimed error exponents.

In this section the packing lemma of \cite{Paris, CAMACEEarXiv} will be generalized in such a way that the two codebooks for the two user are the same. This packing lemma, however, is independent of the channel, so the lemma can be used with any general model from \ref{SubSecGenArbDMC} to get a general exponent derived form the exponent of \cite{Paris, CAMACEEarXiv}.

Achieving a similar result as in e.g. \cite{CAMACEEarXiv}, a constant composition codebook is needed. That means, the codewords $\vx(i)$ (and synch sequence $\vx(0)$), resp. codewords $\vy(j)$ (and $\vy(0)$) are sequences in $\iX^n$ and $\iY^n$, each of the same type $P^X\in\iP^n(\iX)$/$P^Y\in\iP^n(\iX)$ (that means all codewords have the same empirical distribution). If a sequence $\vx$ has type $P^X$ then it is denoted as $\vx\in \iT_{P^X}$.

\begin{figure}[hbt]
\begin{center}
\includegraphics[width=0.4\textwidth]{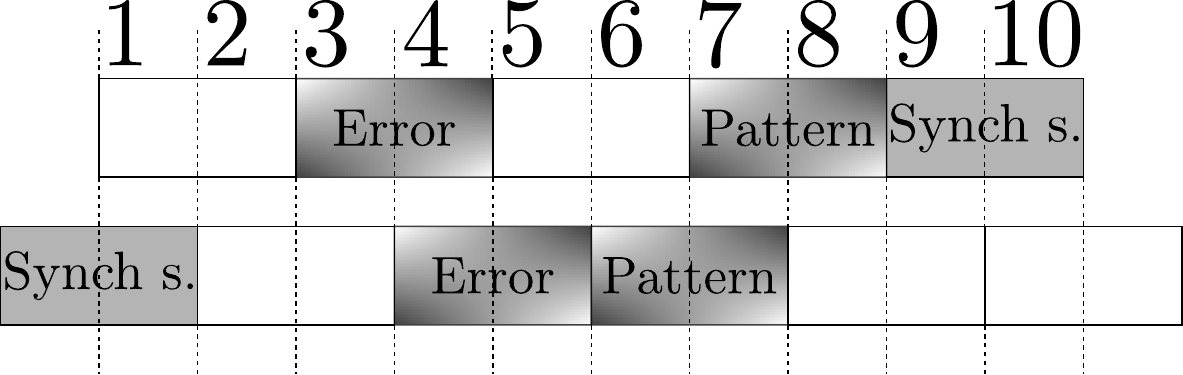}
\caption{ An example for error pattern. Here $S_1=\{3,8\}$, $S_2=\{5,6\}$, $S_{12}=\{4,7\}$.}\label{FigSubblocks}
\end{center}
\end{figure}

To fully comprehend the lemma the concept of \emph{error pattern} must be understand. In asynchronous coding if an error happens then it defines an error pattern, see Figure \ref{FigSubblocks}. Those index of subblocks that are present in the error pattern and corresponding to the first user are denoted by $S_1$, those that are corresponding to the second user are denoted by $S_2$ and those that are corresponding to both user are denoted by $S_{12}$. Let $S=S_1 \cup S_2 \cup S_{12}$, see e.g. \cite{Hawaii} to get more on this. If $\vi, \vj$ decoded as $\hat \vi, \hat \vj$ and the error pattern is $\vS$, the we say $(\vi, \vj, \hat \vi, \hat \vj)\in \iE\iP(\vS)$. 

The next lemma proves the existence of a codebook with ``good'' properties:
\begin{Lem}\label{LemPacking-basic}
For each $K$, $n=2k$, $k\in \mathbb N$ type $P^X\in\iP^k(\iX)$, rate $R<\HH(P^X)-\delta_n,R_2<\HH(P^Y)-\delta_n$ there exists such a codebook $\iC$ with codewords and synch-sequences are from $\iT^k_{P^X}\times \iT^k_{P^X}$ that if it were used as an AMAC code with the two codebooks are $\iC$ and $D=ln+\frac{n}{2}$ then for each error pattern $\vS=(S_1,S_2,S_{12})$ and each $\vV=(V_1, V_2,\dots, V_{2K})$ with $V_{s}\in \iP^{k}(\iX\times\iX\times\iY\times\iY)$, $s\in[2K]$ the following inequality holds:
 \begin{align}
  &\sum_{(\vi,\hat\vi,\vj,\hat\vj) \in \iE\iP(\vS)}\mathds{1}_{V_1, V_2,\dots, V_{2K}}\{\vx(\vi),\vx(\hat\vi),\vy(\vj),\vy(\hat\vj)\}\notag\\
  &\leq p_n\exp\left\{ n(K-1)(R_1+R_2)-\sum\limits_{s\notin S}k\left[\I_{V_s}(X \wedge  Y)\right]\right\}\notag\\ 
  &\quad \cdot\exp\left\{-\sum\limits_{s \in S_1}k\left[\I_{V_s}(\hat X \wedge X\wedge Y)-R_1\right]\right\}\label{EqPack-lem-ineq}\\ 
  &\quad \cdot\exp\left\{-\sum\limits_{s \in S_2} k\left[\I_{V_s}(\hat Y \wedge X\wedge Y)-R_2\right]\right\}\notag\\
  &\quad \cdot\exp\left\{- \sum\limits_{s \in S_{12}} k\left[\I_{V_s}(\hat X \wedge \hat Y \wedge X\wedge Y)-R_1-R_2\right]\right\}\notag
 \end{align}
 where $\mathds 1$ is an indicator function (1 if the sequences has the given subtype sequence) $\vx(\vi)=(\vx(i_1), \vx(i_2), \dots ,\vx(i_K))$, $\delta_n=\frac{3\log n}{n}\max(|\iX|,|\iY|)$   $p_n$ is a polynomial in $n$ that depends only on $|\iX|$ and $K$ and the multi-information defined in \eqref{EqMultiInfDef}.
\end{Lem}
\begin{IEEEproof}[Sketch of the proof\footnote{The original proof of the Lemma in \cite{CAMACEEarXiv} is approximately 6 pages long. Due to the space limitations here, we try to explain how the original proof can be extended to the case where the two codebooks are the same.}]
 The proof will use random selection. Select the codewords from $\iT^k_{P^X}\times \iT^k_{P^X}$ (the first $k$ symbol is from $\iT_{P^X}$ and the last $k$ symbol is also from $\iT_{P^X}$). Now, it must be shown that the expected value of the left side fulfills the statement of the lemma. 
 
 For patterns where all codewords is different (or if all codebook would be different, thus time varying code would be employed) the standard bounds for the type class would provide the desired bounds (see Csiszár and Körner \cite{Csiszar} or \cite{Hawaii,Hongkong}).
 
 The patterns where recurring codewords happen proven by induction on the length of the error pattern, in the same way as in \cite{Paris, CAMACEEarXiv}.
  
 Define auxiliary patterns that are all possible relevant error patterns (without the correctly decoded codewords). For these auxiliary patterns there are analogous bounds defined. If there is no recurrence in an auxiliary pattern then analogous bound can be proven by random selection, in the standard way. If there are recurring codewords, then the standard technique gives an exponential bounds were some of the terms in the exponent are missing. It must be shown that these terms are non-negative so the original upper bounds state.
 
 However, the missing terms define a shorter auxiliary pattern, for which the exponential bound is true (by induction). So, if the missing terms would be negative then the number of codewords that form that pattern would be less than 1. Thus, there would be no codewords that would form that pattern, so their extension to the original pattern with recurring codewords also cannot happen. The number of configurations that cannot happen can be upper bounded by any exponential upper bound. See \cite{Paris, CAMACEEarXiv} for more on auxiliary patterns.
 
 It can be seen, that if the auxiliary patterns are extended to contain the case where the codeword is overlapped with himself then the above technique works, even if the two codebook are the same. As the codewords' first $k$ symbol is independent from the same codeword's last $k$ symbols the statement can be proven in the usual way.
\end{IEEEproof}
\begin{Rem}
 The proof of the achievable exponent can be simplified significantly, as here there is no need to show that the exponent is uniformly continuous with respect to the distributions.
\end{Rem}
\begin{Rem}
 The main practical problem with the Csiszár-type exponents are usually with the polynomial factors. These factors can be dominant in the finite blocklength regime. Our simplified method, however, needs smaller polynomial factor, so the resulting bound may be closer to the practical usage, i.e. for computing the reliability function of the DMC. 
\end{Rem}

\section{Generalizations}\label{SecGeneralization}

\subsection{Extending the result for arbitrary DMC}\label{SubSecGenArbDMC}

\begin{figure}[hbt]
\begin{center}
\includegraphics[width=0.2\textwidth]{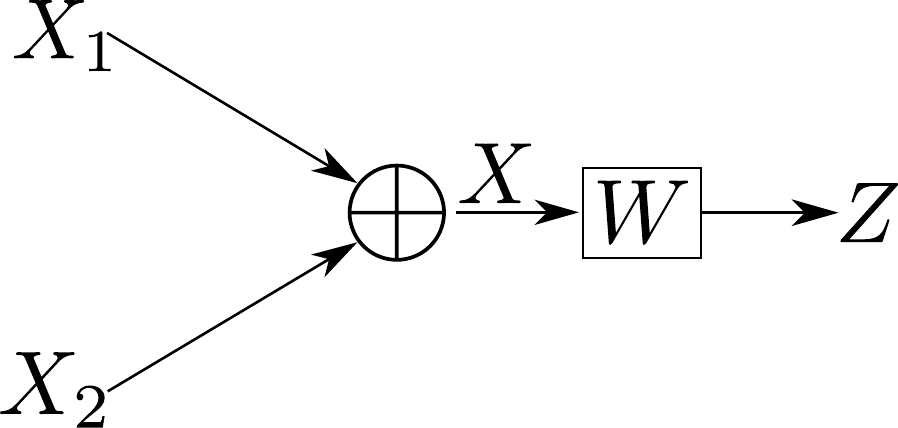}
\end{center}
\caption{ Multiuser channel from single user channel. }\label{FigUjModel}
\end{figure}

Suppose that a single user channel with given input alphabet $\iX$ and output alphabet $\iZ$ is given. The channel matrix $W:\iX \to \iZ$ may be arbitrary and unknown to the sender/receiver.
Suppose further, there is a given binary operation $\iX_1 \oplus \iX_2 \to \iX$. With this operation a virtual MAC can be defined, see Figure \ref{FigUjModel}. The channel matrix $W^{MAC}:\iX_1 \times \iX_2 \to \iZ$ can be derived from $W$ and is unknown to the senders  and the receiver (thus universal coding is required) if $W$ was unknown.

Similarly as in Section \ref{SecKnownResult}, codes for this virtual MAC give rise to trellis codes with memory 1 for the given DMC.

\subsection{Extending the result for memory more than 1}

\begin{figure}[hbt]
\begin{center}
 \includegraphics[width=0.45\textwidth]{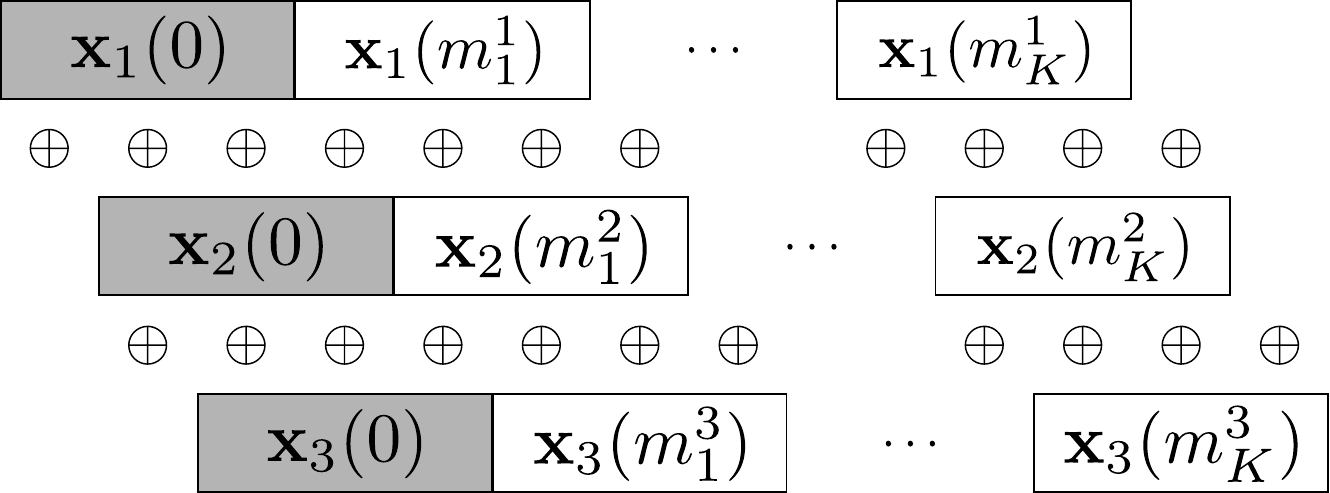}
\end{center}
\caption{ Combining more virtual input streams to a real input stream}\label{FigAsGenModell}
\end{figure}

Suppose we have an $a+1$-ary operation on $\iX$\footnote{Any binary operation can be extended to an $a+1$-ary e.g. \\$x=(\dots((x_1\oplus x_2)\oplus x_3) \dots )\oplus x_{a+1}$}. 
The binary operation $\oplus$ can be extended to $a+1$ inputs. Suppose that the $i$-th sender want to send message sequence $m_1^i, m_2^i, \dots, m_K^i$, and interleave the messages similarly as in Section \ref{SecKnownResult} as $\hat \vm=m_1^1, m_1^2, \dots, m_1^{a+1}, m_2^1, m_2^2, \dots$ (this is the real starting time order of the messages)
then the codewords shifted with $k=\frac{n}{a+1}$ gives a coding system depicted in Figure \ref{FigAsGenModell}
 
Note that, the $s$-th $k$-length input symbol sequence (where counting starts with synch-sequence as 0) 
depends only on $\hat m_{s}, \hat m_{s-1}, \dots, \hat m_{s-a}$. Indeed, this is a Trellis code with memory $a$.

Please note that this paragraph is quite speculative, as no packing lemma or exponent has been proven form more than 2 transmitter yet. However, we do not see any obstacles that would prevent the generalization of Packing lemma and exponent to more than 2 user case. If the Packing lemma and the exponent can be generalized to more senders, then the above method can give exponent for trellis coding system whose memory is more than 1 (say $a$ in this case), time invariant and universal.

\section{Outlook}

With the help of Lemma \ref{LemPacking-basic} and a properly chosen binary operation the existence of time-invariant trellis code of memory 1 is proven from arbitrary DMC. The exponent of this code can be derived from the MAC extension of this DMC (that is given by the binary operator) and the general asynchronous error exponent from \cite{CAMACEEarXiv}.

In the 2 user case and also the multi-user case, it seems also to be possible to slightly modify the decoder, to get a coding decoding system with similar exponent without synch-sequences. The proof of this result needs further investigation. 

It seems the proving technique of the packing lemma can be extended, to get error exponent for a universal, time-invariant trellis coding technique of arbitrary long memory. 

\section*{Acknowledgment}
 I thank Imre Csiszár and Tamás Kói and the unkonwn reviewers for their numerous suggestions, conversation and corrections regards this matter.

\printbibliography

\end{document}